\documentclass[runningheads]{llncs}

\usepackage{fontawesome}  
\usepackage{amsmath}
\usepackage{multirow}
\usepackage[table,xcdraw]{xcolor}
\usepackage{graphicx}
\usepackage{textcomp}
\usepackage{subfig}
\usepackage[utf8x]{inputenc}
\usepackage{csquotes}
\usepackage{flexisym}
\usepackage{eurosym}

\usepackage{lscape}

\newcommand{\ignore}[1]{}
\usepackage[hidelinks]{hyperref}
\usepackage[font=scriptsize]{caption}

\usepackage[ruled,vlined,linesnumbered]{algorithm2e}

\usepackage{amssymb}
\usepackage[bottom]{footmisc}
\usepackage{wrapfig}
\usepackage{nicefrac}
\usepackage{bm}
\usepackage{mathrsfs}
\usepackage{xcolor}
\usepackage{hyperref}
\newcommand{\RN}[1]{%
	\textup{\uppercase\expandafter{\romannumeral#1}}%
}

\usepackage{mathtools}

\newcommand{\orcid}[1]{
	\href{https://orcid.org/#1}{\includegraphics[scale=0.4]{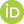}}
}

\begin{document}
\title{Trustworthy Artificial Intelligence and Process Mining: Challenges and Opportunities}
\titlerunning{Trustworthy AI and PM: Challenges and Opportunities}
%
%

\author{Andrew Pery\inst{1}\orcid{0000-0001-6443-0489}\textsuperscript{\href{mailto:andrew.pery@abbyy.com}{\faEnvelopeO}} \and
	Majid Rafiei\inst{2}\orcid{0000-0001-7161-6927} \and
	Michael Simon\inst{3}\orcid{0000-0003-1891-5973} \and
	Wil M.P. van der Aalst\inst{2}\orcid{0000-0002-0955-6940}}

\authorrunning{Andrew Pery et al.}
%
%
\institute{ABBYY, Ottawa, Canada\\
	\and
	Chair of Process and Data Science, RWTH Aachen University, Aachen, Germany\\
	\and
	XPAN Law Partners, Boston, USA
}
\maketitle              
\vspace{-0.5 cm} 
\begin{abstract}

The premise of this paper is that compliance with Trustworthy AI governance best practices and regulatory frameworks is an inherently fragmented process spanning across diverse organizational units, external stakeholders, and systems of record, resulting in process uncertainties and in compliance gaps that may expose organizations to reputational and regulatory risks. Moreover, there are complexities associated with meeting the specific dimensions of Trustworthy AI best practices such as data governance, conformance testing, quality assurance of AI model behaviors, transparency, accountability, and confidentiality requirements. These processes involve multiple steps, hand-offs, re-works, and human-in-the-loop oversight.
In this paper, we demonstrate that process mining can provide a useful framework for gaining fact-based visibility to AI compliance process execution, surfacing compliance bottlenecks, and providing for an automated approach to analyze, remediate and monitor uncertainty in AI regulatory compliance processes.


\end{abstract}
\section{Introduction}\label{sec:introduction}

AI-based technologies are becoming pervasive, impacting virtually every facet of our lives. While AI has a lot of promise, not all of its impacts are good.
There is growing evidence that AI models can embed human and societal biases and deploy them at scale.
As such, the ever-increasing growth of AI highlights the vital importance of balancing AI utility with the fairness of outcomes, thereby engender a culture of trustworthy AI.  
Fairness is the foundation for Trustworthy AI. Intuitively, fairness seems like a simple concept.  
However, it embodies consideration of a number of dimensions, such as trade-offs between algorithmic accuracy versus human values, demographic parity versus policy outcomes and power-focused questions such as who gets to decide what is fair. 

These are vexing challenges for AI developers, policy-makers and consumers alike.  For AI developers, clarity of what constitutes AI fairness is a key consideration given the juxtaposition of ethical, legal, and reputational issues.  For policy-makers and regulators, the challenge is how to promote innovation while protecting consumers from the harmful impacts of AI. For consumers of AI, its about trustworthiness, whether they can rely upon AI outputs to be accurate and transparent, with safeguards in place to protect them from adverse outcomes.  

This paper explores the challenges and opportunities associated with fostering a culture of Trustworthy AI, with particular focus on: (1) The current state of Trustworthy AI, including a survey of key industry and standards organization initiatives with emphasis on the proposed EU Artificial Intelligence Act, (2) The relationship between Trustworthy AI and Responsible Data Science (RDS), and (3) Contribution of trust aware process mining to facilitate a data-driven analytical framework to surface uncertainties, variabilities, and vulnerabilities in Trustworthy AI compliance processes.

 The remainder of the paper is organized as follows. In Section~\ref{sec:trustworthy_ai}, we define the contours of Trustworthy AI principles. In Section~\ref{sec:eu_proposal}, we explore the proposed EU Artificial Intelligence Act (AIA) that intends to operationalize and implement rigorous risk-based prescriptive processes for ensuring a culture of Trustworthy AI. In Section~\ref{sec:RDS_AI}, we map the relationship between RDS and Trustworthy AI, including a discussion of challenges associated with contextualizing AI fairness as a foundation for Trustworthy AI. In Section~\ref{sec:process_mining}, we discuss the applications and benefits of process mining as an important tool to enable organizations to make data-driven decisions relating to the obligations and conformance requirements inherent in the proposed EU AI regulation.
 

\section{Trustworthy AI}\label{sec:trustworthy_ai}
Surveys reveal an undercurrent of pervasive distrust of AI systems. Cathy O’Neil, a leading advocate for AI algorithmic fairness, highlighted three main reasons behind consumer distrust of AI: \textit{opacity}, \textit{scale}, and \textit{damage} \cite{o2016weapons}.
Fairness is the foundation for trustworthy AI. It is the connective tissue that binds together the principles of ethical use, interpretability, transparency, accountability, and confidentiality that engenders trust and promotes the use of AI for social good. 
Trustworthy AI is a governance framework designed to mitigate potential adverse impacts on consumers as AI is poised to profoundly and indelibly change our lives. 
As mentioned in \cite{trust_AI}, Trustworthy AI is changing the dynamic between user and system into a relationship.


\subsection{Achieving Trust in AI}
Trustworthy AI starts with human agency and autonomy. Trust in AI systems is enhanced when there is a human-in-the-loop who monitors the overall performance of AI systems and when circumstances dictate, remediates potential adverse outcomes. 
Trust in AI is strengthened by giving users the ability to make informed decisions about the impact of AI on their personal and economic well-being.


AI is perceived by consumers to be \textit{a black box}. Data inputs to the AI systems, their learning models, and how they arrive at decisions are neither visible, nor understood by consumers. 
Furthermore, many AI developers defensively protect their algorithms as proprietary and a competitive differentiator.
\textit{Interpretability} and \textit{explainability} of AI are two important elements that strengthen trust in AI.
Interpretability of AI provides insight into the cause and effect between inputs and outputs of an AI system and how AI predicts outcomes.  
Explainability of AI goes one step further by providing users with not only insight into how AI models work but also traceability of AI decisions and documentation relating to the process of data gathering, labeling, and methods used for training AI algorithms.

Consumers have limited recourse to hold AI developers accountable for the adverse impacts of AI systems. While there is sectoral legislation, e.g., Section 5 of the FTC (Federal Trade Commission) Act\footnote{\scriptsize\url{https://www.federalreserve.gov/boarddocs/supmanual/cch/ftca.pdf}}, available for consumers to remedy disparate treatment attributable to AI systems it is an onerous process to prevail. Moreover, for the disparate impact, the burden of proof requires statistical analysis that a protected class is treated differently from others, which is hardly something that would be accessible to average consumers. For these reasons,  accountability, including redress mechanisms in the event of demonstrated harmful impact need to be addressed to achieve trust in AI.  


\subsection{The Emergence of Trustworthy AI Principles}
We can see efforts being made, to varying degrees, that recognize and deal with issues relating to trust in AI by the data sciences community (see Section 4), standards organizations, e.g., IEEE \cite{ieee_standard}, NIST (National Institute of Standards and Technology) \cite{trust_nist}, and by public sector organizations.

In 2019, OECD member countries adopted OECD Council Recommendation on Artificial Intelligence\footnote{\scriptsize\url{https://legalinstruments.oecd.org/en/instruments/OECD-LEGAL-0449}} consisting of five principles of human centered values of fairness of AI, inclusive investments in AI, transparency, accountability, and robustness of AI systems. 
The OECD recommendations were subsequently endorsed by the G20 with particular reference to the view that the \enquote{digital society must be built on trust among all stakeholders including governments, civil society, international organizations, academics, and businesses through sharing common values and principles including equality, justice, transparency, and accountability taking into account the global economy and interoperability}.

While Trustworthy AI principles serve as a helpful framework, they are just that. Adherence to Trustworthy AI is fragmented at best and they lack effective enforcement mechanisms to safeguard against potentially harmful impacts.  For this reason, the momentum has shifted towards the regulation of AI: \enquote{The calls for modest regulation that lets industry take the lead are part of a failed regulatory philosophy, one that saw its natural experiment over the past several decades come up lacking. AI is too important and too promising to be governed in a hands-off fashion, waiting for problems to develop and then trying to fix them after the fact}.\footnote{\scriptsize\url{https://www.brookings.edu/research/ai-needs-more-regulation-not-less/}}

\section{The Proposed EU Regulation of AI}\label{sec:eu_proposal}

On April 20, 2021 the European Commission released the proposal for the regulation of artificial intelligence\footnote{\scriptsize\url{https://ec.europa.eu/commission/presscorner/detail/en/ip_21_1682}}, the ambition of which is to balance the socio-economic benefits of AI and new risks or negative consequences for individuals or society.   
The proposed Artificial Intelligence Act (AIA) takes a risk-based approach to regulate AI by fostering an \enquote{ecosystem of trust that should give citizens the confidence to take up AI applications and give companies and public organisations the legal certainty to innovate using AI}. 
In the following, we demonstrate five governing principles for trustworthy AI proposed by AIA.


\subsection{Scope of the Proposed Regulation}
The proposed AIA applies to all providers, i.e., natural or legal persons, public authorities, agencies, or any other body that develops an AI system, that places or makes available on the market or puts into service AI systems or services in the EU (cf. Article 3).
The AIA also assigns responsibility to users, importers, distributors, and operators who make use of or make substantial modifications to the functionality and performance of AI systems (cf. Article 26-29).   
The geographic scope for the AIA will operate irrespective of whether such providers are established in the EU or a third country, and so will cover where the system users are in the EU or the output of the systems is used in the EU (cf. Article 2).     
AI systems under the regulation encompass a wide range of methods and algorithms including supervised, unsupervised, and reinforcement machine learning for a given set of human-defined objectives that generate outputs such as content, predictions, recommendations, or decisions influencing the environments they interact with (cf. Article 3).     

\subsection{Risk-Based Approach}
The foundation of the AIA is a risk-based approach that classifies AI systems into three categories based on a combination of factors that include the intended purpose, the number of impacted persons, and the potential risk of harms (cf. Article 5-7): 
\begin{itemize}
    \item Prohibited AI: Systems that use subliminal techniques that cause physiological or psychological harm, exploit vulnerable groups, effectuate social scoring by public authorities that may result in discrimination or unfavorable treatment, and remote biometric systems used by law enforcement in public spaces (subject to well-defined exceptions) (cf. Article 5).
    \item High Risk: Annex III  provides a list of systems that are used in critical infrastructures, educational or vocational training, human resources, essential private and public services, law enforcement, migration, asylum and border control management, and administration of justice and democratic processes (cf. Article 7).
    \item Low Risk: While not explicitly named (we use the term \textit{low risk} of our own choosing), by default, all systems not categorized as \textit{prohibited} or \textit{high-risk}. Providers of such systems are encouraged to institute responsible use of AI best practices on a voluntary basis (cf. Article 69).
\end{itemize}

\subsection{Promote Fair and Trustworthy AI Best Practices}
The AIA sets forth a comprehensive legislative mandate to ensure fairness in the application of AI systems that safeguards fundamental human values and promotes socio-economic rights. Some of these mandates are as follows: obligation on providers to implement appropriate risk management measures throughout the entire lifecycle of AI systems (cf. Article 9), rigorous data governance processes (cf. Article 10), technical documentation, and record-keeping processes to enable monitoring of compliance (cf. Article 11-12), transparency that enables full interpretation of outputs (cf. Article 13), and Human-in-the-loop oversight (cf. Article 14).

\subsection{Transparency and Accountability}
According to the AIA, providers of AI systems will be required to implement a range of processes to ensure full transparency into and accountability for AI systems (cf. Article 19-23) such as (1) conformity assessment and certification processes, (2) auditability, including accessible event logs, and (3) Explainability, potentially to coordinate with the human-in-the-loop for adjudication and remediation.

\subsection{Enforcement}
The AIA incorporates an onerous enforcement mechanism that even surpasses the fines under the GDPR (cf. Article 71). Some examples are as follows: up to €10m or 2\% of the total worldwide annual turnover for the supply of incorrect, incomplete or misleading information to the authorities, up to €20m or 4\% of the total worldwide annual turnover for non-compliance with any other AIA requirement or obligation, and up to €30m or 6\% of the total worldwide annual turnover for violations of prohibited practices. 

While the proposed AIA is far from ratification and still subject to vigorous debate within the EU Parliament and Council, the momentum towards its adoption is inevitable.
Like the GDPR, the AIA will serve as a model for other jurisdictions that will seek to finally exert control over what has been the unregulated, hyperbolic growth of AI across the globe.

\section{Responsible Data Science and Trustworthy AI}\label{sec:RDS_AI}
Responsible Data Science (RDS) is a discipline that is influential in shaping Trustworthy AI best practices.  
RDS refers to the collection of techniques and approaches trying to reap the benefits of data science and big data while ensuring \textit{fairness}, \textit{accuracy}, \textit{confidentiality} and \textit{transparency} \cite{van2016responsible}.
To minimize adverse AI outcomes of AI the role of RDS is to: (1) Avoid unfair conclusions even if they are true, i.e., the fairness principle, (2) Answer questions with a guaranteed level of accuracy, i.e., the accuracy principle, (3) Answer questions without revealing secrets, i.e., the confidentiality principle, and (4) Clarify Answers such that they become indisputable, i.e., the transparency principle.

\begin{figure}[tb]
    \centering
    \includegraphics[width=\columnwidth]{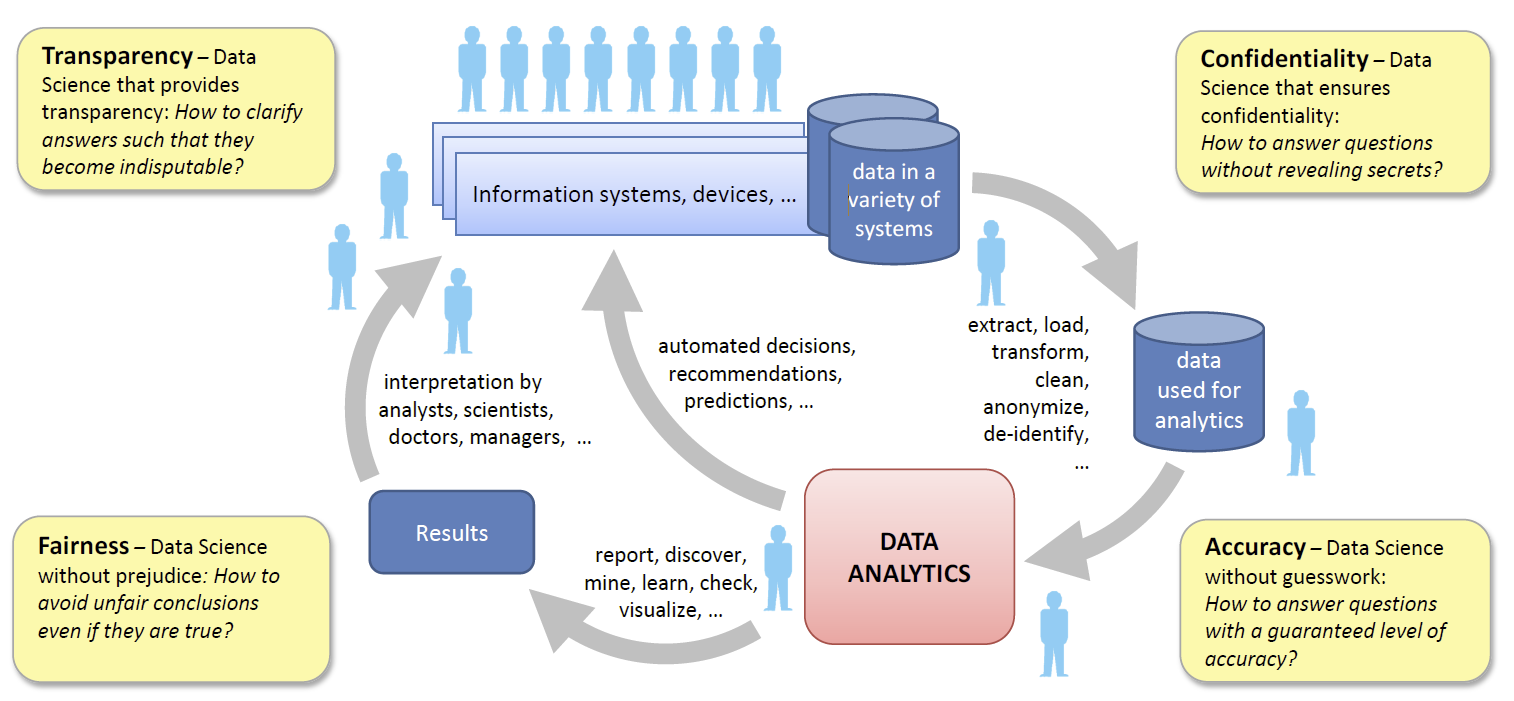}
    \caption{The data science pipeline facing the four FACT challenges \cite{van2016responsible}.}
    \label{fig:RDS}
\end{figure}

RDS applies a methodology throughout the entire life cycle of information to support trustworthy AI best practices by applying these four principles of fairness, accuracy, confidentiality, and transparency to the \textit{data science pipeline} resulting in rigorous data governance as illustrated in Figure~\ref{fig:RDS}.

RDS delivers a robust framework for the ethical design of AI systems that addresses the following key areas: (1) Unbiased outcomes through the application of appropriate fairness constraints to the training data, (2) Algorithmic outcomes interpreted in a manner that is meaningful to end users, (3) Resilience in how AI systems deliver accurate results and respond to change in inputs, (4) Accountability for the system’s outcomes, and (5) Safeguarding the confidentiality of training data through privacy enhancing measures.
However, providing each aspect of RDS has its own challenges from contextualizing the aspect to implementing it in data science and AI systems. In \cite{challenges_conf}, the authors describe the challenges regarding the \textit{confidentiality} aspect for \textit{process mining} which combines process and data science. In the following, we provide the challenges regarding the \textit{fairness} aspect. 

\subsection{Contextualizing Fairness in AI Systems: Challenges}\label{subsec:fairness_in_AI} 
The idea of \textit{fairness} is somewhat amorphous.
At its highest level of abstraction, fairness is a normative concept that comes from our conscience. Dator defines a fair system as follows: \enquote{What is fairness then? We all have desires and we want people to treat us according to those desires. We also know that people around us have similar desires and want to be treated accordingly. Fairness is closely related to fair play so it seems logical to conclude that a fair system is a system where everybody is treated in a similar way} \cite{fairness_book}.
There are a number of challenges associated with contextualizing and applying such a high level of abstraction to a more concrete algorithmic AI fairness framework.

First, fairness may be influenced by cultural, sociological, economic, and legal considerations. What may be considered as fair in one culture may be perceived as unfair in another. Unequal distribution of opportunity may require the application of distributive fairness that levels the playing field. 
For example, in the context of credit applications, there ought to be an equal probability of loan eligibility by ensuring that AI algorithmic outcomes do not discriminate against members of protected groups \cite{group_fairness}. There are other instances where the application of corrective fairness may be necessary, for example, to remedy adverse impacts in the administration of justice, housing, education, and employment.


Second, equality does not necessarily result in the fairness of outcomes. While under Human Rights legislations disparate treatment on the basis of race, gender, nationality, disability, and sexual orientation is prohibited there may still be instances of adverse outcomes, based on other facially-neutral variables that cause a disparate impact, i.e., unintentional discrimination \cite{fairness_dwork}.
Consider Amazon’s free same day delivery service based on an AI algorithm that included attributes, such as distance to the nearest fulfillment center, local demand in designated zip code areas, and frequency distribution of prime members to determine profitable locations for free Same-Day Delivery.  
The algorithm was found to be biased against minorities even though race was deemed not to be a factor in the determination of same day delivery, and minority residents in the selected zip codes were \textit{about half as likely} to be eligible as white residents.\footnote{\scriptsize https://eu.usatoday.com/}


The third challenge is balancing algorithmic fairness with fairness outcomes \cite{algorithm_fariness}.
In this context, fairness encompasses policy and legal considerations, and leads us to ask: \textit{what ought to be fair?} For example, in the context of hiring practices, what ought to be a fair percentage of women in management positions that AI algorithms should incorporate as thresholds to promote gender parity? 

The fourth challenge relates to trade-off in balancing demographic parity with the utility of outcomes. For example, if AI algorithms remove disparate impact in the incarceration of minorities, how would that impact broader policy considerations such as the administration of justice? 

Finally, fairness implicates issues of power.
Before we can decide what is fair, we need to decide who gets to decide that. The conundrum we must confront is that the minority groups who are so typically the victims of algorithmic bias are rarely given a seat at the table when it is time to define what is fair. The unfortunate result is that far too often, the definition of fairness is simply what those already in power need it to be to maintain that power.  
 
\subsection{Implementing Fairness: Challenges for Data Scientists}
Fairness constraints need to be considered in the context of specific use cases and for desired outcomes. Bias may be introduced at various levels within an AI system. Training data may introduce proxies that discriminate. Historical bias may unconsciously result in adverse outcomes, for example through word embeddings \cite{fairness_book}. Representation bias through under or, over representation of training data may produce disparate impacts. The algorithms may not sufficiently adjust for fairness constraints. Inadequate testing for disparate treatment and impact may have adverse consequences for protected groups.
While some argue that AI algorithms in fact minimize bias there is compelling evidence that they can and often amplify biases.  Examples span facial recognition, criminal justice, hiring practices, and loan approvals \cite{250171}. 

Regardless of any contextualization, any definition, and any implementation approach of the fairness which is the cornerstone for Trustworthy AI, what is essential is to gain visibility to and remediate potential gaps in Trustworthy AI compliance processes. In the next section, we demonstrate how process mining could play a role in fulfilling such requirements.

\section{Process Mining for Promoting Trustworthy AI }\label{sec:process_mining}
Compliance with the proposed EU AIA requires an understanding of process execution and interactions between multiple internal and external stakeholders, risk assessment of diverse systems of record that incorporate AI systems, and cooperation with various regulatory bodies and standards organizations.

The proposed AI regulation operationalizes and codifies trustworthy AI principles with prescribed mandates to institute \textit{appropriate data governance and management practices}. The governance mechanism is complex and requires human and systems-based interactions between diverse internal and external stakeholders and EU and national regulators.
Monitoring conformance with AIA is delegated to national supervisory authorities, they are empowered to order companies to take corrective actions, access all information, documentation, and data required to enforce compliance with the proposed regulation. 

Given the complexity and variability of interactions implicit in achieving compliance with the proposed regulation it is our contention that \textit{process mining} can be a valuable tool to help organizations gain visibility to various dimensions of prescribed process flows stipulated by the regulation, accelerate the analysis of how information flows, surface process bottlenecks, visualize interactions generated by event logs from disparate systems of record that may reveal areas of compliance and reputational risks. 
Process mining bridges the gap between data science and process science using event data captured from different types of information systems \cite{van2016process}.
It is a data-driven approach that enables organizations to gain insight into interactions between people, systems, and organizations based on “as-is" visualization of process execution.

There are many techniques and activities in the context of process mining. However, the three main types of activities in process mining are \textit{process discovery}, \textit{conformance checking}, and \textit{enhancement}. Process discovery techniques take an event log and discover a process model without using any other information. Conformance checking techniques take a process model and an event log of the same process to check whether reality, as recorded in the event log, conforms to the model and vice versa. Enhancement techniques are used to extend or improve a given process model using the information about the process recorded in some event logs \cite{van2016process}.

Process Mining can facilitate compliance with AIA by many functionalities such as: (1) Surfacing AI regulatory compliance process gaps and uncertainties, (2) Capturing user interactions performing compliance tasks, (3) Comparing process execution variations, (4) Highlighting compliance task outliers and errors, (5) Identifying potential root causes for improper execution, (6) Real-time monitoring of processes to ensure conformance to prescribed process execution paths, and (7) Triggering alerts in the event of non-compliant process tasks or changes in conditions. 
Furthermore, the AIA proposed regulation is inherently collaborative in nature wherein process execution spans across different organizations.

As discussed in \cite{trust_mining}, in collaborative processes where different organizations execute different parts of a shared process, the internal activities carried out by each organization are beyond the control of the other collaborators resulting in uncertainty regarding process execution. Whenever there is uncertainty in a process, there is a need for trust. Hence, collaborative business processes are especially trust-intensive. In such trust-intensive environments, process mining can be used to clarify the flow of activity execution among several organizations.




\begin{figure}[]
    \centering
    \includegraphics[width=0.99\columnwidth]{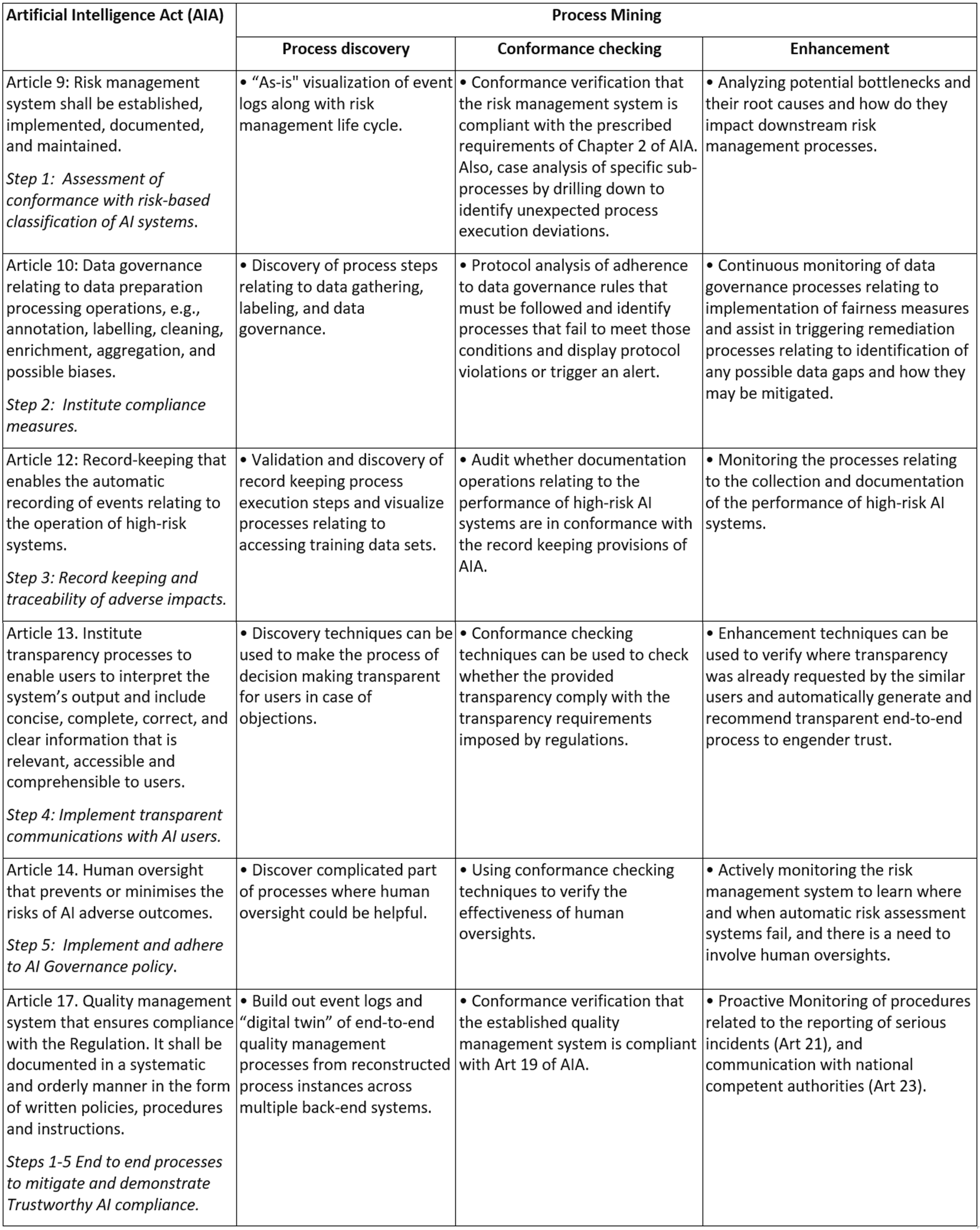}
    \caption{Process mining cadence to meet AIA prescriptive compliance obligations.}
    \label{fig:mapping}
\end{figure}

Compliance with AIA constitutes a number of interdependent steps. Performing these steps may involve variabilities in process execution paths and hand off between different stakeholders and prescribed conformance obligations to meet Articles 16-23 and Annex VII of the AIA:

\begin{itemize}
    \item Step 1: R\&D teams develop and bring to market AI systems in accordance with the risk classification system defined by the proposed regulation. If it is a high-risk AI system then a priori conformance assessment must be undertaken and a declaration of conformity must be submitted to the appropriate National Supervisory Authority. Then the AI system may be placed on the market.
    \item Step 2: Legal and Compliance teams must institute compliance measures in accordance with Chapter 2 of the proposed regulation that ensures adherence to data governance, accountability, transparency, accuracy, robustness, and cybersecurity provisions. 
    \item Step 3: Data Science teams must undertake continuous monitoring of AI systems, collect data on the system’s operation and take corrective action if needed. The post-market monitoring system must actively and systematically collect, document, and analyze relevant data provided by users.
    \item Step 4: Customer-facing functions such as Sales, Marketing, and Support, are responsible for providing clarity and certainty as to the expected AI system inputs and outputs in a way that users are informed that they are interacting with an AI system, augmented with human oversight who monitor their operation and be able to decide, to override or reverse the output of the high-risk AI system.
    \item Step 5: Implementation of a Quality Management System with auditable and traceable documentation relating to the techniques, procedures for the design, of the high-risk AI systems, including procedures for data management, data analysis, data labeling, data storage, data aggregation, data retention and report serious incidents that may result in adverse outcomes.  
\end{itemize}

Figure~\ref{fig:mapping} further maps the compliance steps, the obligation provisions of the AIA, and process mining functionality to support Trustworthy AI. The figure illustrates how process mining techniques can facilitate AIA obligations. The FACT challenges of RDS are also taken into consideration in process mining as a subdiscipline called Responsible Process Mining (RPM) which is recently receiving increasing attention \cite{rafiei_CEDP_arxiv,rafiei_group_elsevier,Gamal_mine_me,pripel_short}.


\section{Conclusion}\label{sec:conclusion}
Trustworthy AI engenders a climate of trust essential for achieving sustainable competitive advantages in an intensely competitive environment where the application of AI is a disruptive force.
The proposed EU regulation of AI is a comprehensive prescriptive measure which imposes onerous obligations, redress mechanisms on AI developers and businesses deploying AI systems.
To mitigate compliance, reputational, and business risks process mining is poised to provide a data-driven approach to discover how existing Trustworthy AI compliance processes work, surface and remediate process bottlenecks, visualize different pathways of process execution and identify and remediate variations from prescribed protocols. Process mining can be a useful toolbox for ensuring that certain AI systems are designed and developed in accordance with common necessary requirements before they are put on the market and operationalized through harmonized technical standards. 
    

\section*{Acknowledgments} Funded under the Excellence Strategy of the Federal Government and the L{\"a}nder. We also thank the Alexander von Humboldt Stiftung for supporting our research.

\bibliographystyle{splncs04}
\bibliography{Refrences}

\end{document}